\documentclass[prb,twocolumn,groupaddress]{revtex4}
\usepackage{mleftright}
\usepackage{graphicx}
\usepackage{tabularx}
\usepackage{capt-of}
\usepackage{color}
\usepackage{subfigure}
\usepackage{multirow}
\usepackage{array}
\usepackage{amsmath,amssymb,amsfonts,textcomp}
\usepackage{color}
\usepackage{subfigure}
\usepackage{multirow}
\usepackage{array}
\usepackage{physics}
\usepackage{booktabs}
\usepackage{array}
\newcolumntype{P}[1]{>{\centering\arraybackslash}p{#1}}
\usepackage{hyperref}

\ExplSyntaxOn
\NewDocumentCommand{\longdash}{ O{2} }
 {
  --\prg_replicate:nn { #1 - 1 } { \negthinspace -- }
 }
\ExplSyntaxOff

\newcommand{\thickhline}{%
    \noalign {\ifnum 0=`}\fi \hrule height 1pt
    \futurelet \reserved@a \@xhline
}

\begin{document}

\begin{abstract}
  Bootstrap embedding (BE) is a recently developed electronic structure method that has shown great success at treating electron correlation in molecules. Here, we extend BE to treat surfaces and solids where the wave function is represented in periodic boundary conditions using reciprocal space sums (i.e. $k$-point sampling). The major benefit of this approach is that the resulting fragment Hamiltonians carry no explicit dependence on the reciprocal space sums, allowing one to apply traditional non-periodic electronic structure codes to the fragments even though the entire system requires careful consideration of periodic boundary conditions. Using coupled cluster singles and doubles (CCSD) as an example 
method to solve the fragment Hamiltonians,
we present minimal basis set CCSD-in-HF results on 1D conducting polymers.
We show that periodic BE-CCSD can typically recover
$\sim$99.9\% of the electron correlation energy.
We further demonstrate that periodic BE-CCSD is feasible even for complex donor-acceptor polymers of interest to organic solar cells - despite the fact that the monomers are sufficiently large that even a $\Gamma-$point periodic CCSD calculation is prohibitive. We conclude that BE is a promising new tool for applying molecular electronic structure tools to solids and interfaces.
\end{abstract}

\title{Periodic Bootstrap Embedding}

\author{Oinam Romesh Meitei}
\affiliation{Department of Chemistry, Massachusetts Institute of Technology, Cambridge, Massachusetts 02139, United States}

\author{Troy Van Voorhis}
\email{tvan@mit.edu}
\affiliation{Department of Chemistry, Massachusetts Institute of Technology, Cambridge, Massachusetts 02139, United States}

\maketitle
\section{Introduction}

The lack of affordable, accurate electronic structure methods limits computational insights into solids and surfaces. Due to their enormous (in principle, infinite) size, the electronic structure of solids and interfaces must be treated under periodic boundary conditions. To efficiently treat these systems, one must further exploit the translational symmetry of the system by representing the wave function in reciprocal space. Currently, only a handful of methods, such as density functional theory (DFT), dynamical mean-field theory (DMFT), or random phase approximation (RPA), are routinely used to perform periodic computations on extended systems\cite{https://doi.org/10.1002/jcc.20080,doi:10.1146/annurev.physchem.59.032607.093528, https://doi.org/10.1002/jcc.540140608, RevModPhys.68.13, doi:10.1080/00268976.2011.614282}. 

On the other hand, many wavefunction-based electronic structure methods exist for non-periodic systems and are readily available in standard quantum chemistry codes. Methods such as second-order Moller-Plesset perturbation (MP2), truncated coupled cluster (CCSD), complete active space (CAS), or the selected configuration interaction (SCI), have been used with great success to predict a variety of electronic, structural, and thermodynamic properties of molecules and supramolecular complexes. Historically, only a few of these approaches have been adapted for computation under the periodic boundary condition, for example, MP2 and CCSD\cite{gusJCP2001,bartlettCPL2001,bartlettJCP2004,eomCCJCP2005,gruneisJCTC2011,berkelbachJCTC2017}. Much ongoing work is devoted to extending a greater variety of molecular electronic structure tools to solids and making existing tools more efficient.

Embedding based-methods have shown promising results in adapting the aperiodic electronic structure methods to treat periodic systems\cite{PhysRevB.76.045107,PhysRevB.98.085138, https://doi.org/10.1002/qua.25801, doi:10.1063/1.4903828, doi:10.1021/acs.jctc.8b00927,doi:10.1063/5.0084040, doi:10.1063/1.469264, doi:10.1021/acs.jctc.0c00576, doi:10.1021/acs.jctc.6b00651, https://doi.org/10.1002/wcms.1357,PhysRevB.20.5345, INGLESFIELD200189}. Notable works include density functional embedding theory, projection-based wavefunction-in-DFT methods, and density matrix embedding theory (DMET)\cite{GOVIND1998129, CarterJCP2002,GoodpasterJCTC2018, doi:10.1021/acs.jctc.9b00571, GusDMETsolid2014, ChanDMETsolid2020, chanDMETDMFT2020, GagliDMETsolid2020}. In these approaches, only the sub-systems are treated with electron correlation methods while the bulk of the system under the periodic boundary condition is treated at a mean-field level of theory.

Recently, we developed a fragment-based quantum embedding method called Bootstrap Embedding (BE) to treat electron correlation for molecules and supramolecular systems\cite{beJCTC2020,beJCTC2019, beJPCL2019, beJCP2020}. Unlike other existing embedding approaches, BE provides scopes for flexible system partitioning using fragments with overlapping regions. The method utilizes matching conditions of the wavefunction in the overlap regions to improve the embedding. Numeical tests on various molecular systems have demonstrated the approach's accuracy and applicability. 

Herein, we implement BE to treat electron correlation under periodic boundary conditions for infinite solids and surfaces. We show that traditional, aperiodic correlated methods can be interfaced with BE to treat periodic systems. Starting with a mean-field solution at the thermodynamic limit (TDL), BE converges the correlated calculation to the TDL without resorting to reciprocal space summation. We demonstrate that BE provides an inexpensive approach to treating electron correlation at the TDL by computing total energies for 1D conducting polymers.
The periodic implementation of BE typically recovers $\sim$99.9 \% of the total electron correlation energy with CCSD as the method of choice (BE-CCSD). We further demonstrate the applicability of periodic BE to complex polymers with large unit cell sizes.

This work is organized as follows: Section \ref{sec:EmbHam} provides the formulation of the embedding Hamiltonian under periodic boundary conditions. Section \ref{sec:Fragments} discusses the fragment schemes and Section \ref{sec:PBE} describes periodic BE. Section \ref{computational} outlines the computational details of periodic BE calculations. The performance and accuracy of periodic BE are presented in Sections \ref{sec:conv} and \ref{sec:accuracy} compared to full CCSD results with $k$-point sampling. Section \ref{sec:psc} demonstrates the applicability range of periodic BE by performing computation on two complex donor-acceptor polymers. Finally, Section \ref{conclusion} summarizes the main results of this work.

\section{Theory}\label{theory}

In this section, we briefly review Bootstrap Embedding (BE) and outline the main framework of BE for periodic systems. For a detailed discussion on the general formalism and machinery of BE, we refer to Ref. \citenum{beJCTC2020}.

\subsection{Periodic Embedding Hamiltonian}\label{sec:EmbHam}

Our approach for defining the embedding Hamiltonian largely follows Ref. \citenum{ChanDMETsolid2020} and \citenum{GagliDMETsolid2020}. We present the formalism here to lay the groundwork for defining fragments and matching conditions in Sections \ref{sec:Fragments} and \ref{sec:PBE}, respectively. In usual practice, mean-field computations under periodic boundary conditions are performed on a grid of finite points or $k$-points in the reciprocal-space first Brillouin zone (FBZ). Thus, the resulting Hartree Fock (HF) solution has $k$ sets of molecular orbitals corresponding to solutions on each of the finite $k$-points on the grid. We can analogously define atom-centered local orbitals (LO) for each $k$-point.

\begin{align}
  \psi^k_n(r) = \sum_{\mu} C^k_{\mu n}\chi^k_{\mu}(r)
\end{align}

The LO defined by the above equation can be constructed, for example, using L\"{o}wdin symmetric orthogonalization (the method of choice used in this work). Other techniques include intrinsic atomic orbitals\cite{iaoGerald2013} or maximally localized Wannier functions\cite{wannierPRB1997}.

A discreet Fourier transform relates $k$-point representation and the real-space representation. For convenience, we transform the $k$-point LO to a real-space representation.

\begin{align}
  \psi^{\bf R}_n(r) = \frac{1}{\sqrt{N_k}}\sum_k e^{-ik{\bf R}}\psi^k_n\label{eq:k2r}
 \end{align}
  
Hereafter, we omit the {\bf R}-index when dealing with a real-space representation.

In BE, a subset of LOs, $N_A$, is defined as fragment $A$. The choices of fragments are discussed in the next section.
The fragment orbitals are chosen as the subset of LOs centered on a given subset of atoms - with the choice of atoms again being described in the next section.

Following a Schmidt decomposition, the HF state can be distinctly partitioned into\cite{geraldDMET2013}:

\begin{align}
  \ket{\phi_{HF}} = \left(\sum_{p=1}^{N_A} \lambda^A\ket{f^A_p}\otimes\ket{b^A_p}\right)
    \otimes  
         \ket{\Phi^{env,A}} \label{eq:schmidth}
\end{align}

In the above equation, we have further partitioned the non-fragment space into bath states, $\ket{b^A}$, entangled with the fragment state, $\ket{f^A}$, and the environment, $\ket{\Phi^{env,A}}$, disentangled from the fragment. This definition considerably reduces the Hilbert space of the total system.

In practice, the Schmidt decomposition described above is achieved by singular value decomposition (SVD) of the off-diagonal HF density matrix in the LO basis between the fragment and remainder sites.

\begin{align}
P^{A} = U_A \Sigma_A V^{\dagger}_A
\end{align}\label{Eq:svd}

Working in the real-space representation, $P^A$ by definition has a dimension of $(N_{total}-N_A)\times N_A$. $N_{total}$ is the total number of LO in the real-space representation.

The left-singular vectors, $U_A$, define the bath orbitals in Equation \ref{eq:schmidth}. We can then construct a transformation to the embedding basis comprising the fragment and the bath orbitals as below.

\begin{align}
T^{{\bf R},A}=\mleft[
\begin{array}{c|c}
  I &  \\
  \hline
   & U_A
\end{array}
\mright]
\end{align}

The embedding Hamiltonian for each fragment is then constructed by transforming the molecular Hamiltonian using $T^{{\bf R},A}$. Since the one-electron and two-electron integrals are usually computed in the $k$-point representation under the periodic boundary condition, $T^{{\bf R},A}$ absorbs a phase factor $e^{-ik{\bf R}}$ and is denoted by $T^{k,A}$ below.

\begin{align}
\hat{H}^A = \sum^{2N_A}_{pq}h^A_{pq}a^{\dagger}_p a_q + \frac{1}{2}\sum^{2N_A}_{pqrs}V^A_{pqrs}a^{\dagger}_pa^{\dagger}_ra_sa_q\label{Eq:frag_hamiltonian}
\end{align}

where 
\begin{align}
h^A_{pq} = \frac{1}{N_k}\sum_k \sum_{\mu\nu}^N T^{k,A}_{\mu p} F^k_{\mu\nu} T^{k,A}_{\nu q} - V^{HF,A}_{pq}\label{Eq:oneelec}
\end{align}
\begin{align}
V^A_{pqrs} = \frac{1}{N_k}\sum_k \sum_{\mu\nu\lambda\sigma}^N T^{k,A}_{\mu p}T^{k,A}_{\nu q}V^k_{\mu\nu\lambda\sigma}T^{k,A}_{\lambda r}T^{k,A}_{\sigma s}\label{Eq:eri}
\end{align}

where $F^k$ is the Fock matrix, $V^k$ is the two-electron integrals, and $V^{HF,A}$ is the HF potential constructed using the HF density matrix in the embedding basis, $P^{HF,A}$.

Equations \ref{Eq:oneelec} and \ref{Eq:eri} show that the resulting fragment Hamiltonian $H^A$ has \textit{no explicit dependence on either $k$-space or real space indices outside the fragment or bath}. Instead, those dependencies are wrapped up in the definition of new effective interactions, $h^A$, and $V^A$, within the fragment and bath space.
The consequence is that any non-periodic method can be used to solve for the ground state of the fragment Hamiltonian. In particular, existing non-periodic code can be used without significant modification as a fragment solver for a periodic system. Further, because only the Hamiltonian (and not the correlated calculation) depends on $k$-points, it is very inexpensive to approach the thermodynamic limit (TDL) in an embedding calculation; one effectively only needs to converge the mean field calculation to the TDL and the embedded calculation will also be converged.

The integral transformation of the two-electron repulsion integrals (ERI) in Equation \Ref{Eq:eri} from the atomic orbital basis to the embedding basis is one of the computationally demanding steps in BE. To reduce computational cost in the transformation, we have utilized density fitting with gaussian-type orbitals as the choice of auxiliary basis functions. Details on the ERI transformation in the presence of $k$-points can be found in Ref. \citenum{ChanDMETsolid2020}.

The ERI transformation in Equation \Ref{Eq:eri} contains redundant computations as two or more fragments share the same set of atom-quartet ERI in an overlapping region. The redundant computation can be remedied by first computing the integral transformation for all unique atom-quartets and then recovering the fragment ERI for each fragment from the set of the unique atom-quartets. Ref. \citenum{beJCTC2020} discusses this algorithm in more detail. The minimal basis calculations presented in Section \ref{sec:results} does not require the efficient integral transform and, therefore, was not implemented. However, we anticipate that significant computational savings would be possible for more complex systems.

\subsection{Fragment Construction}\label{sec:Fragments}
As mentioned earlier, a subset of LOs defines a fragment in the BE framework. The subset of LOs centered on the same atom is considered the minimal unit. Following our earlier work, each atom in the unit-cell defines a fragment composed of LOs on the atom and all atoms connected up to $(n - 1)$ coordination shell. With $n=1$, each atom in the unit cell is a fragment of its own. The nearest connected neighboring atoms are included in the fragments for $n=2$, while $n=3$ also includes the second nearest neighboring atoms. Here forth, BE$n$ denotes this scheme. Such a fragment definition provides scope for using larger fragments in the BE framework.

\begin{figure}
\includegraphics[scale=0.7]{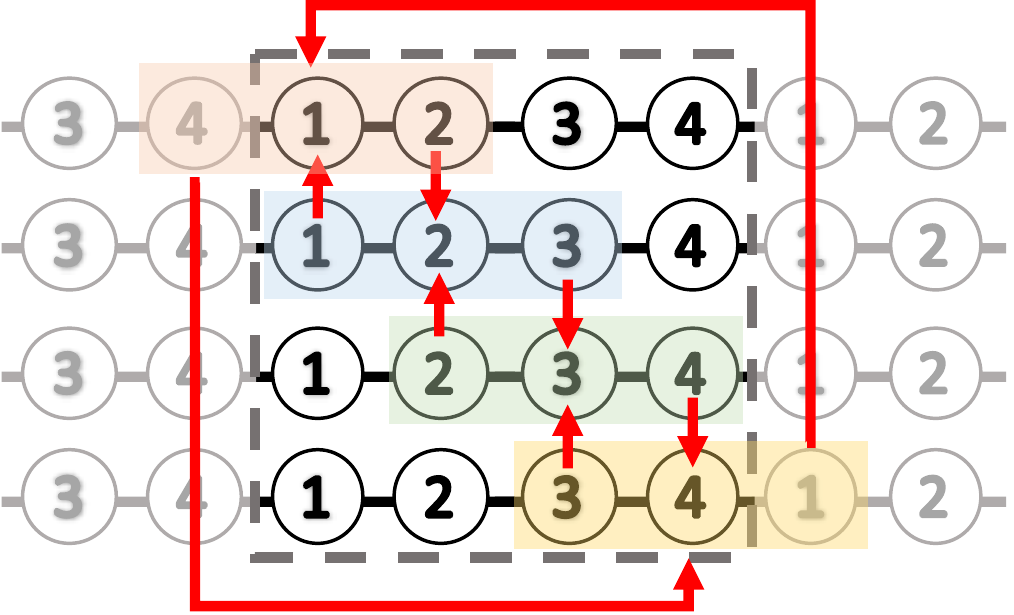}
\caption{Illustration of fragments used in bootstrap embedding. The four fragments have distinct center and edge sites. Each site in the unitcell (gray box) is the center of a fragment. The density matching conditions for each fragents are shown in red arrows.}\label{Fig:be_fragment}
\end{figure}

Considering periodicity, the fragment scheme considered in this work also uses the atom-centered LOs of the neighboring unit cells. There are as many fragments as atoms in the unit cell. The resulting fragments have overlapping regions, with a set of $N_A^{\mathbb{E}}$ LOs appearing as an edge on fragment $A$ and as a center on at least another fragment $B\neq A$. Figure \Ref{Fig:be_fragment} illustrates fragments in the BE2 scheme for a unit cell with four atoms.

\subsection{Periodic Bootstrap Embedding}\label{sec:PBE}

Central to BE, the wave function on edge LOs in one fragment, $\mathbb{E}_A$, is matched to the corresponding wave function on the center LOs in another fragment, $\mathbb{C}_B$, where it overlaps in terms of a one-electron reduced density matrix (1-RDM). The embedding Hamiltonian (Equation \Ref{Eq:frag_hamiltonian}) is solved at a high-level method such as CCSD (the method used in this work) to obtain the 1-RDM. The main idea is that the wave function on each fragment's center is more accurately described in the high-level computation than the wave function on edge. The reason is that the edges interact more strongly with the corresponding mean-field bath orbitals. The problem then boils down to the following constrained optimization:

\begin{align}
\min_{\Psi^A} \bra{\Psi^A}\hat{H}^A\ket{\Psi^A}
\end{align}
subject to
\begin{align}
P^A_{pq} = P^B_{pq}, \quad\forall p,q \in \mathbb{E}_A \cap \mathbb{C}_B, \quad\forall B\ne A\label{Eq:match_condition}
\end{align}

where $P^A_{pq} = \bra{\Psi^A} a_q^{A\dagger}a_p^A\ket{\Psi_A}$ is the 1-RDM. We loop over all fragments, B for which the edge sites in fragment A, $\mathbb{E}_A$, appear as the center sites, $\mathbb{C}_B$. The exact formulation can be derived for all other possible fragments. Further, a global constraint is imposed on the center of each fragment to preserve the total electron count.

The above-constrained optimizations lead to a set of eigenvalue equations for each fragment. Here, the embedding Hamiltonian is dressed with a fragment-specific effective potential, $\lambda_{pq}^A$, that ensures the density matching conditions and global chemical potential, $\mu$, that fixes the electron count. The resulting eigenvalue equation is, then,

\begin{align}
\left( \hat{H}^A + \sum_{pq\in\mathbb{E}_A} \lambda_{pq}^A a^{A\dagger}_p a_q^A + \mu\sum_{p\in\mathbb{C}_A} a^{A\dagger}_p a_p^A \right)\ket{\Psi^A} \nonumber \\
= \mathcal{E^A}\ket{\Psi^A}
\end{align}

The above eigenvalue equation is solved using a high-level electron correlation method in a self-consistent fashion until Equation \Ref{Eq:match_condition} is satisfied.

Following Ref. \citenum{Booth2022}, the total BE energy is computed as:

\begin{align}
  E = E_{HF}^{[0]} + \sum^{N_{frag}}_A\sum_{p\in\mathbb{C}_A}\Bigg[\sum_q^{2N_A}F_{pq}^{A,[0]}\Delta P^A_{pq} \nonumber\\
    + \frac{1}{2}\sum^{2N_A}_{qrs}V^A_{pqrs}K^A_{pqrs}\Bigg]\label{Eq:energy}
\end{align}

where $E^{[0]}_{HF}$ is the reference HF energy, $F^{[0]}$ is the Fock matrix corresponding to the reference HF density in the embedding basis, $\Delta P^A$ is the difference in the correlated 1-RDM, $P^A$ and the reference HF density, $P^{HF, A}$ of the fragment, and $K^A$ is an approximate two-body cumulant defined in terms of the true two-body cumulant of the fragment, $\tilde{K}$, as:

\begin{align}
K_{pqrs}^A = \tilde{K}_{pqrs}^A + \Delta P_{pq}^A\Delta P_{rs}^A - \frac{1}{2}\Delta P_{pr}^A \Delta P_{sq}^A
\end{align}

The two-electron reduced density matrix, $\Gamma^A$ and the true two-body cumulant, $\tilde{K}^A$ are related by:

\begin{align}
\Gamma_{pqrs}^A = P^A_{pq}P^A_{rs} - \frac{1}{2}P^A_{pr}P^A_{sq} + \tilde{K}_{pqrs}
\end{align}

As discussed in Ref. \citenum{Booth2022}, we note that using the approximate two-body cumulant, $K^A$, avoids constructing the Fock matrix, $F^{A,[P^A]}$ corresponding to the correlated 1-RDM, $P^A$.
We note that the cumulant-based energy in Equation \ref{Eq:energy} provides \emph{dramatically} better correlation energies for the systems in this paper as compared to the original density-matrix-based energy used in, for example, Refs. \citenum{GusDMETsolid2014, chanDMETDMFT2020, beJCTC2020, iaoGerald2013, geraldDMET2013}. In some cases, large correlation energy errors are reduced to less than 1\% by the cumulant-based approach, in line with the observations in Ref. \citenum{Booth2022}.
  
\section{Computational Details}\label{computational}

The accuracy of periodic BE was tested by computing the total electron correlation energy per unit cell for 1D polyethylene (PE), polyacetylene (PA), poly(p-phenylene) (PPP), poly(p-phenylene vinylene) (PPV), and polythiophene (PT) chains. BE calculations were also performed on a modified structure of the NT-812 and the PM6 polymers. The side alkane chains on these polymers were removed and replaced by a hydrogen atom. All the structures, including the lattice constants, are provided in Supporting Information. STO-3G basis set was used throughout. The BE energies were compared to full $k$-point CCSD (denoted $k$-CCSD hereafter) correlation energies. The $k$-CCSD calculations were performed with a Monkhorst-Pack $k$-point sampling that included up to $(1 \times 1 \times 18)$ $k$-point grids.
The $k$-CCSD computation for extremely dense $k$-points is challenging even for the relatively small systems (PE and PA) considered in this work. For BE calculations, periodic Hartree-Fock solutions with $k$-points sampling up to $(1 \times 1 \times 45)$ $k$-point grids were used.

\begin{figure}
\includegraphics[scale=0.39]{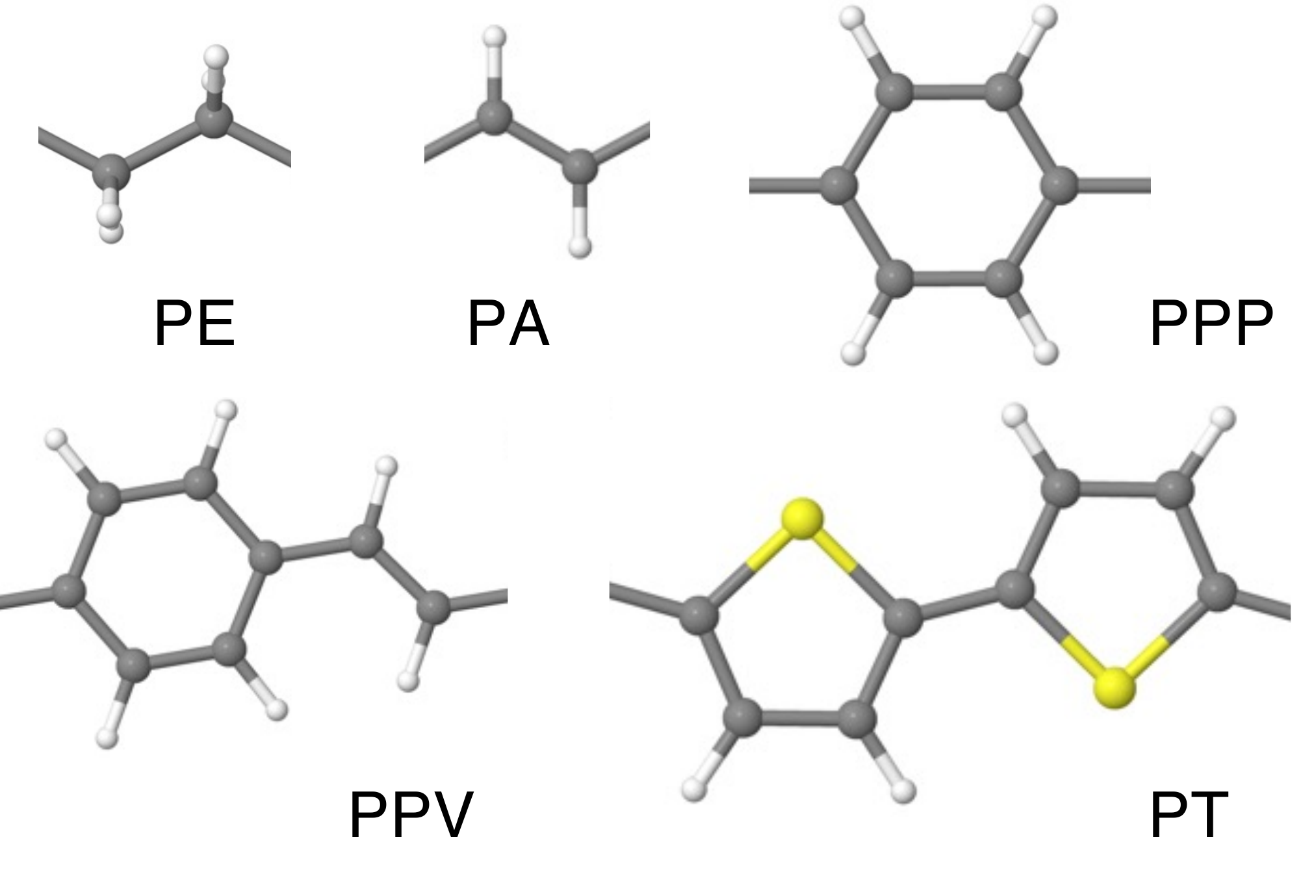}
\caption{Unit cell of the various polymers used in this work.}\label{Fig:small_geom}
\end{figure}

Following our earlier work, an $unrelaxed$ CCSD one-electron reduced density matrix (1-RDM) was used for the matching condition in Equation \ref{Eq:match_condition}. The total electron correlation energy per unit cell obtained from BE, as well as $k$-CCSD calculations, are extrapolated to the thermodynamic limit (TDL) using a power law expansion of the form\cite{eomccSolidBerkelbach2020}:

\begin{align}
E(N_k) = E_{\infty} + \alpha N^{-1}_k + \beta N^{-2}_k
\end{align}

The BE computations were performed with an in-house code using PySCF to generate the necessary integrals in Equations \ref{Eq:oneelec} and \ref{Eq:eri} and for the CCSD solver (non-periodic implementation)\cite{pyscf2018,pyscf2020}. A Quasi Newton-Raphson optimization algorithm was employed to perform the BE optimization with approximate Jacobian as described in Ref \citenum{beJCTC2020}. The BE matching conditions were satisfied when the threshold set for the root-mean-square in the difference of 1-RDM for all the fragments is below $1\times10^{-6}$.

\section{Results and Discussions}\label{sec:results}
\subsection{Convergence with $k$-points}\label{sec:conv}

\begin{figure}[htb!]
\begin{minipage}{0.5\textwidth}
\includegraphics[scale=0.35]{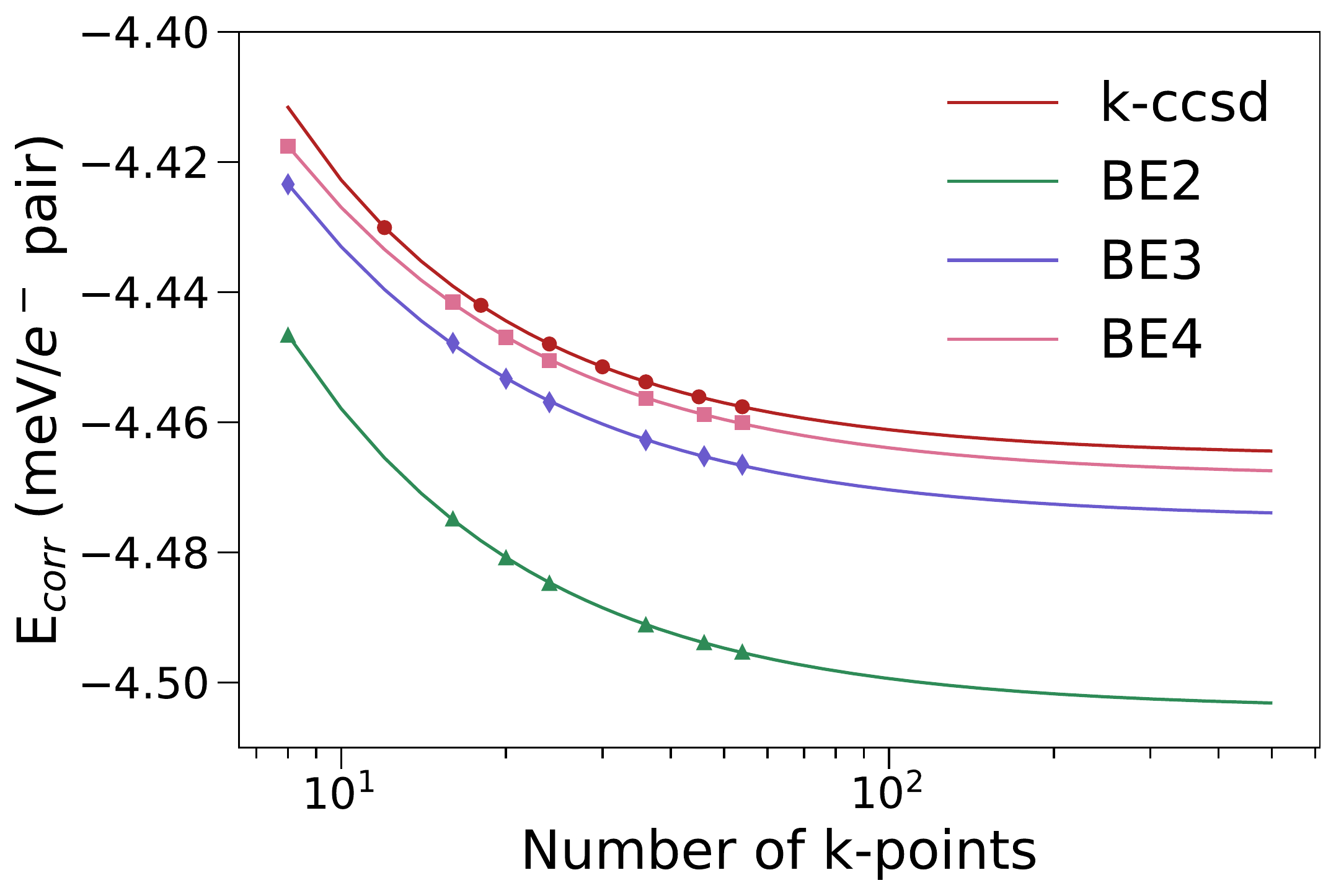}
\end{minipage}
\caption{Convergence of the total electron correlation energy per unit cell, rescaled per electron pair, to the thermodynamic limit for 1D polyacetylene chain.}\label{Fig:PE_conv}
\end{figure}

This section discusses the performance of periodic BE compared to full $k$-CCSD calculations. In particular, we look into the convergence of total electron correlation energy as a function of $k$-points used to sample the first Brillouin zone (FBZ). Figure \ref{Fig:PE_conv} presents the convergence of the total electron correlation energy per unit cell to the TDL for the polyacetylene (PA) chain from various BE$n$ schemes and full $k$-CCSD, rescaled per electron pair. 1D PA chains are widely used in literature to benchmark the accuracy of new methods\cite{bartlettJCP2004, eomCCJCP2005}. Notably, the convergence to the TDL for the PA chain strongly depends on the density of $k$-points used to sample the FBZ. With increasing system size, the correlation strength substantially changes in the PA chain; a considerable $k$-point computation is necessary to converge the electron correlation energy to the TDL. Convergence plots for the other polymer chains considered in this work are provided in the Supporting Information.

BE$n$ clearly reproduces the convergence of $k-$CCSD correlation energies to TDL with an absolute error of 
39, 10, and 3 $\mu$eV per electron pair for BE2, BE3, and BE4, respectively, at TDL. The BE$n$ correlation energy also converges with respect to the fragment size, going from BE2 with smaller fragments to BE4 with larger fragments at each $k-$points. The same is true for the other polymer chains considered in this work; for the polyethylene (PE) chain, BE4 almost exactly reproduces the $k-$CCSD correlation energy (see Supporting Information).
The main distinction to the k-point CCSD computation is that correlation computation in BE$n$ calculation has no explicit dependence on the number of $k$-points. And thus, the computational cost of BE$n$ calculation depends rather weakly on the reciprocal space mesh sizes. On the other hand, the computational cost of the full $k$-CCSD increases rapidly and becomes intractable. 
Given the relatively small sizes of the 1D polymers considered in this work, we can still converge the total electron correlation energies with the full $k$-CCSD for reference. It is, however, different for extended-sized unit cells, such as the polymers in Section \ref{sec:psc}, where even a $\Gamma$-point full $k$-CCSD computation is not feasible.

\begin{figure}[htb!]
\includegraphics[scale=0.35]{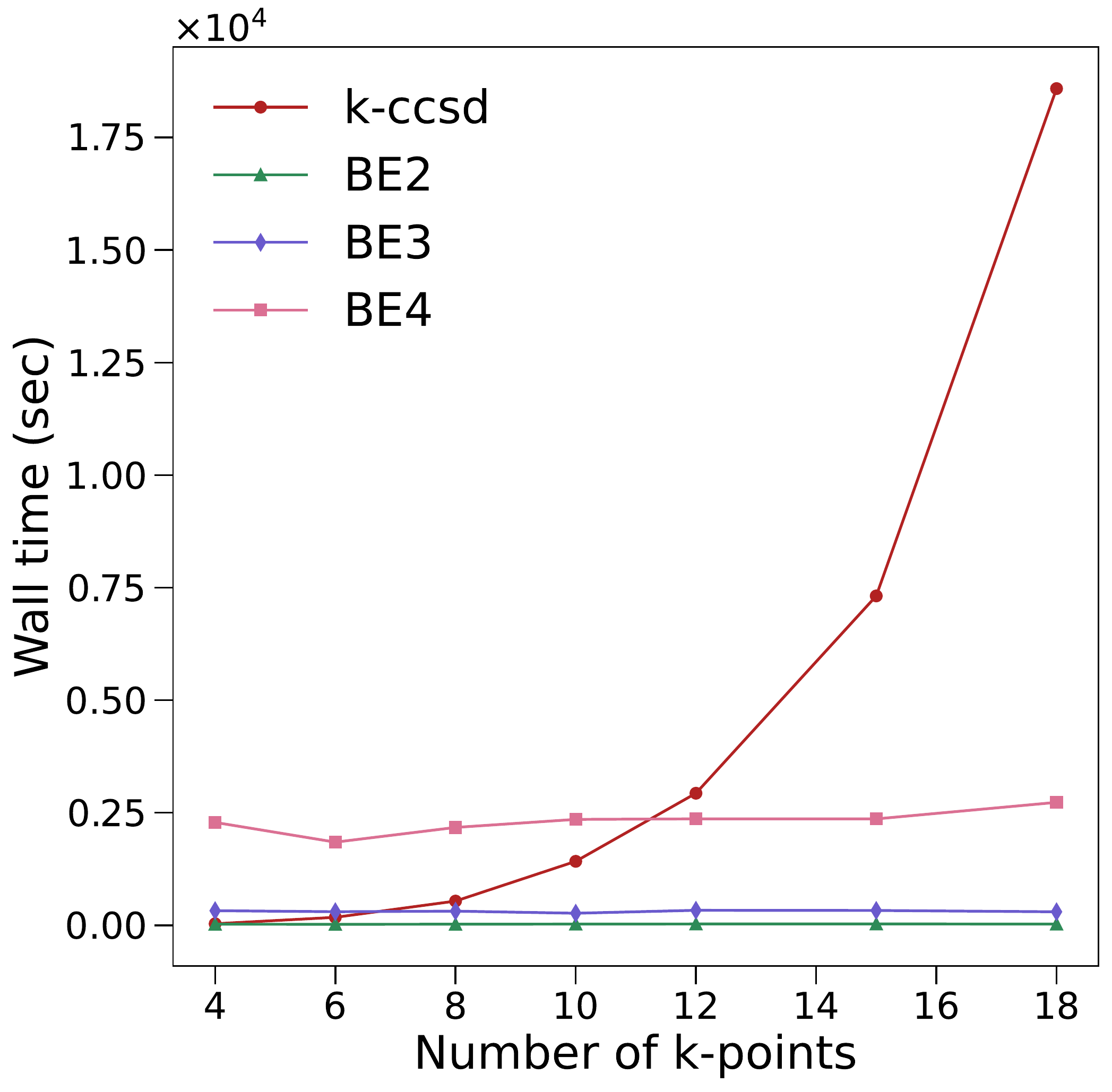}
\caption{Computational time for k-CCSD, BE2, BE3, and BE4 with respect to the number of k-points of polyacetylene chain.}\label{Fig:walltime}
\end{figure}

Figure \ref{Fig:walltime} presents the total wall time required for computing the total electron correlation energies with the BE$n$ schemes and the full $k$-CCSD calculations as a function of the number of $k$-points for the PA chain. The computational cost of BE$n$ hardly depends on the number of $k$-points, whereas the full $k$-CCSD scales quadratically with the reciprocal space mesh sizes.
Although the computational cost of BE2 is lower than the full $k$-CCSD for all the mesh sizes, BE3 and BE4 calculations become less expensive for mesh sizes large than 6 and 10 $k$-points, respectively. However, as visible in Figure \ref{Fig:PE_conv}, these crossing points are far from converging the electron correlation energies to the TDL. Thus, with BE$n$, the electron correlation energy at extremely dense k-points can be accessed at almost the same cost of computations on fewer k-points, guaranteeing convergence to the TDL. 
It is also noteworthy that BE2 and BE3 have relatively similar computational costs, whereas BE4 is comparatively more expensive than the two BE schemes.

\subsection{Accuracy of Periodic BE}\label{sec:accuracy}

\begin{figure}[htb!]
\includegraphics[scale=0.35]{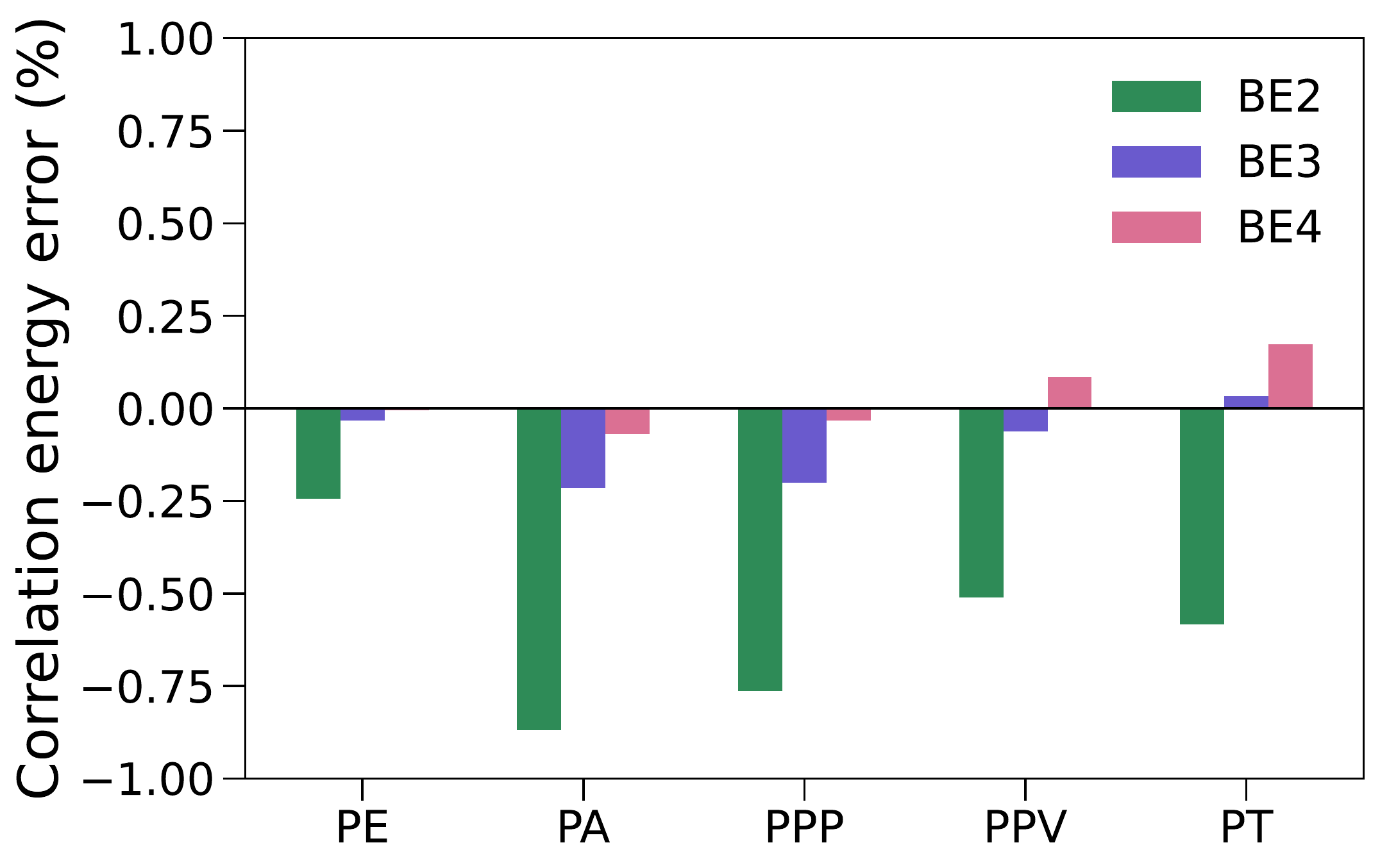}
\caption{Percentage error at the thermodynamic limit. The total electron correlation energy per unit-cell from the BE2, BE3 and BE4 are compared to the full k-point CCSD correlation energy.}\label{Fig:tdl_error}
\end{figure}

To establish the accuracy of periodic BE, we now compare the total electron correlation energies at the TDL computed from the BE$n$ schemes to the full $k$-CCSD method for PE, PA, PPP, PPV, and PT polymer chains. PPP, PPV, and PT are conducting polymers with many applications, such as light-emitting diodes and photovoltaic devices\cite{PPPPPVgeom, PTgeom}. Full $k$-CCSD computations with dense reciprocal space mesh sizes are very challenging on these polymers due to large memory requirements. Figure \ref{Fig:tdl_error} presents the percentage error in total electron correlation energy per unit cell at the TDL from the various BE$n$ schemes compared to full $k$-CCSD energies.
The mean absolute errors in the total electron correlation energies per unit cell at TDL for the polymer chains are 0.59\%, 0.11\%, and 0.07\% with BE2, BE3, and BE4, respectively.

BE2 underestimates the full $k$-CCSD correlation energy with an error between 0.3\% and 0.9\% in the polymer chains. BE3, on the other hand, yields correlation energy with an error between 0.03\% and 0.21\% as compared to the full $k$-CCSD correlation energy. BE4, with the largest fragment, has an error between 0.005\% and 0.173\%.
The systematic errors are in the order of less than 0.04, 0.01, 0.004 meV per electron pair with BE2, BE3, and BE4, respectively (see Supporting Information).
As also seen in the previous section, the BE electron correlation converges to the $k-$CCSD correlation energies with an increase in fragment size for all the polymers. A slight deviation is observed in the PT chain, where BE3 has an error of 0.03\% while BE4 has an error of 0.17\%.
We note that some of this may be due to the difficutly of converging the reciprocal space sums for $k$-CCSD in this case; compared to PE, PPP and PPV, PT is much more sensitive to the number of k-points. Noting that an error of only 12 $\mu$eV per electron pair would be sufficient to make BE4 more accurate than BE3 in this case, it is plausible that more accurate $k$-CCSD calculations might reverse this situation.
BE3 and BE4 yield highly accurate correlation energies for the polymer chains considered in this work, typically recovering about 99.9\% of the total electron correlation energy per unit cell.
Out of the three BE schemes, BE3 best balances the tradeoff between accuracy and the computational time. The total electron correlation energies at various $k$-points from the BE$n$ schemes, as well as full $k$-CCSD calculations for the polymers, are provided in Supplementary Information.

\begin{figure}[htb!]
\includegraphics[scale=0.5]{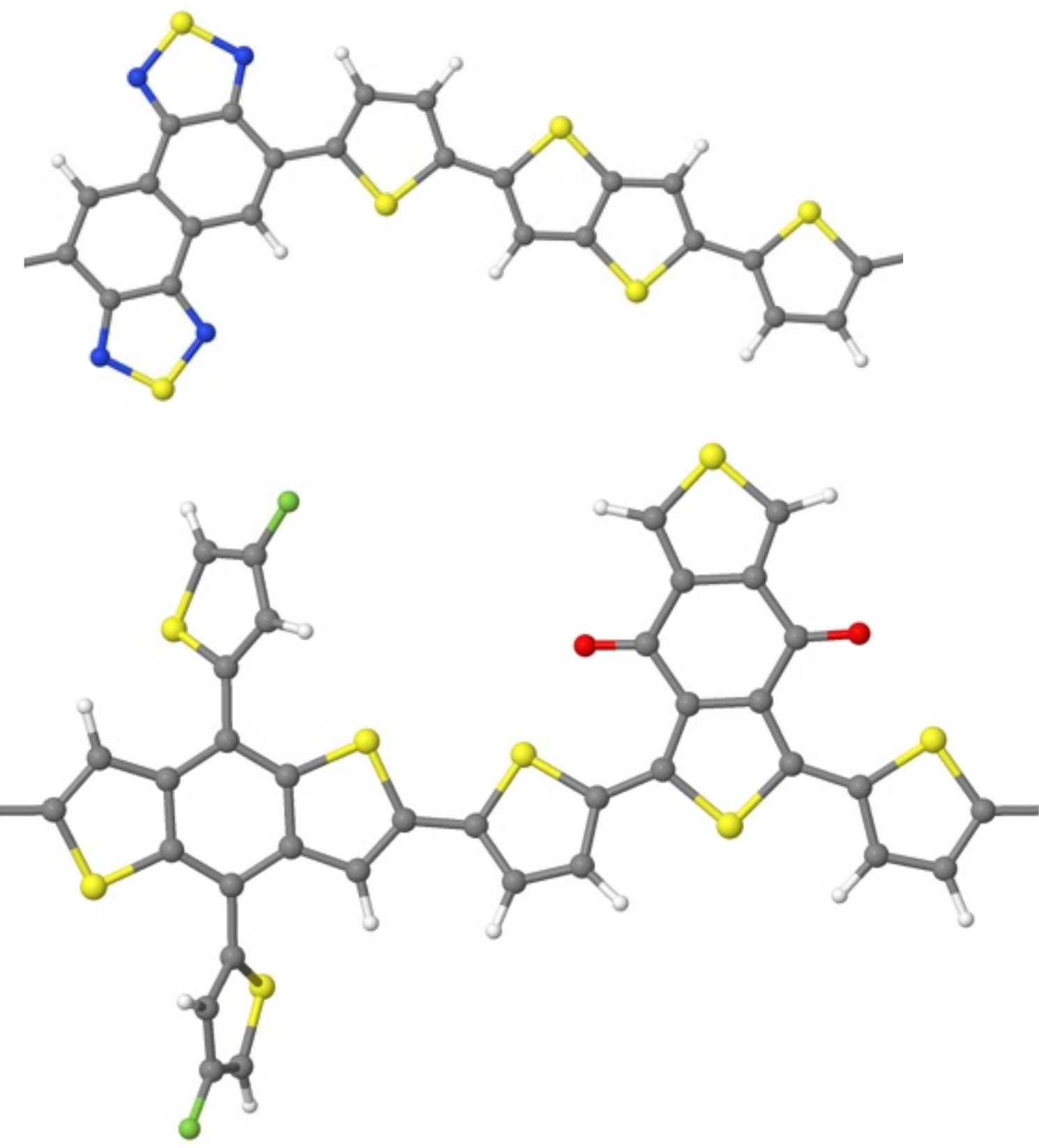}
\caption{Unit cell for the modified NT812 (upper) and PM6 (lowe) polymers. The side alkane chains in the original structures have been replaced with a hydrogen atom.}\label{Fig:large_geom}
\end{figure}

\subsection{Application to Polymer Solar Cells}\label{sec:psc}

\begin{figure}[htb!]
\begin{minipage}{0.5\textwidth}
(a)\\
\includegraphics[scale=0.35]{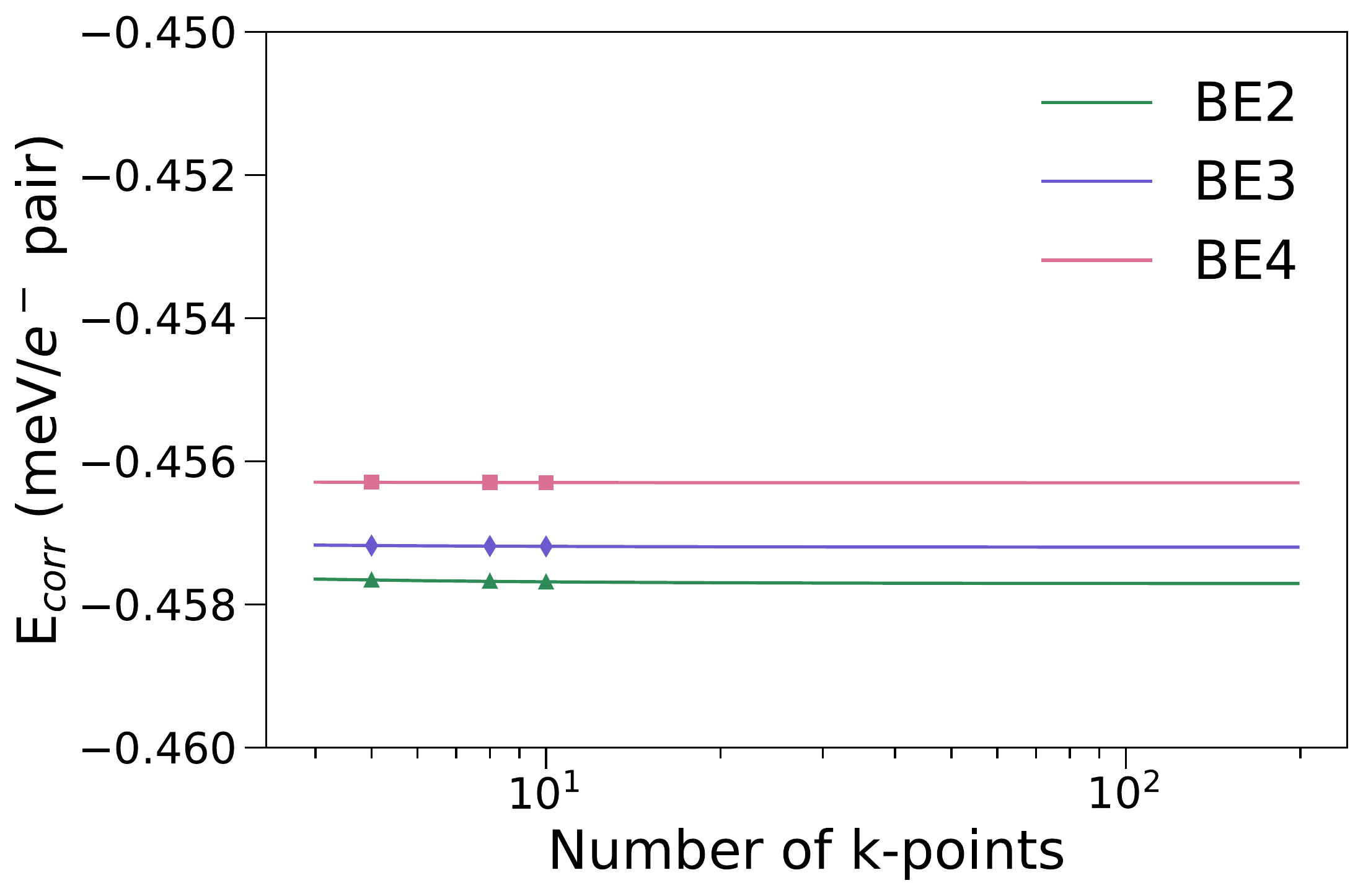}
\end{minipage}\\
\begin{minipage}{0.5\textwidth}
(b)\\
\includegraphics[scale=0.35]{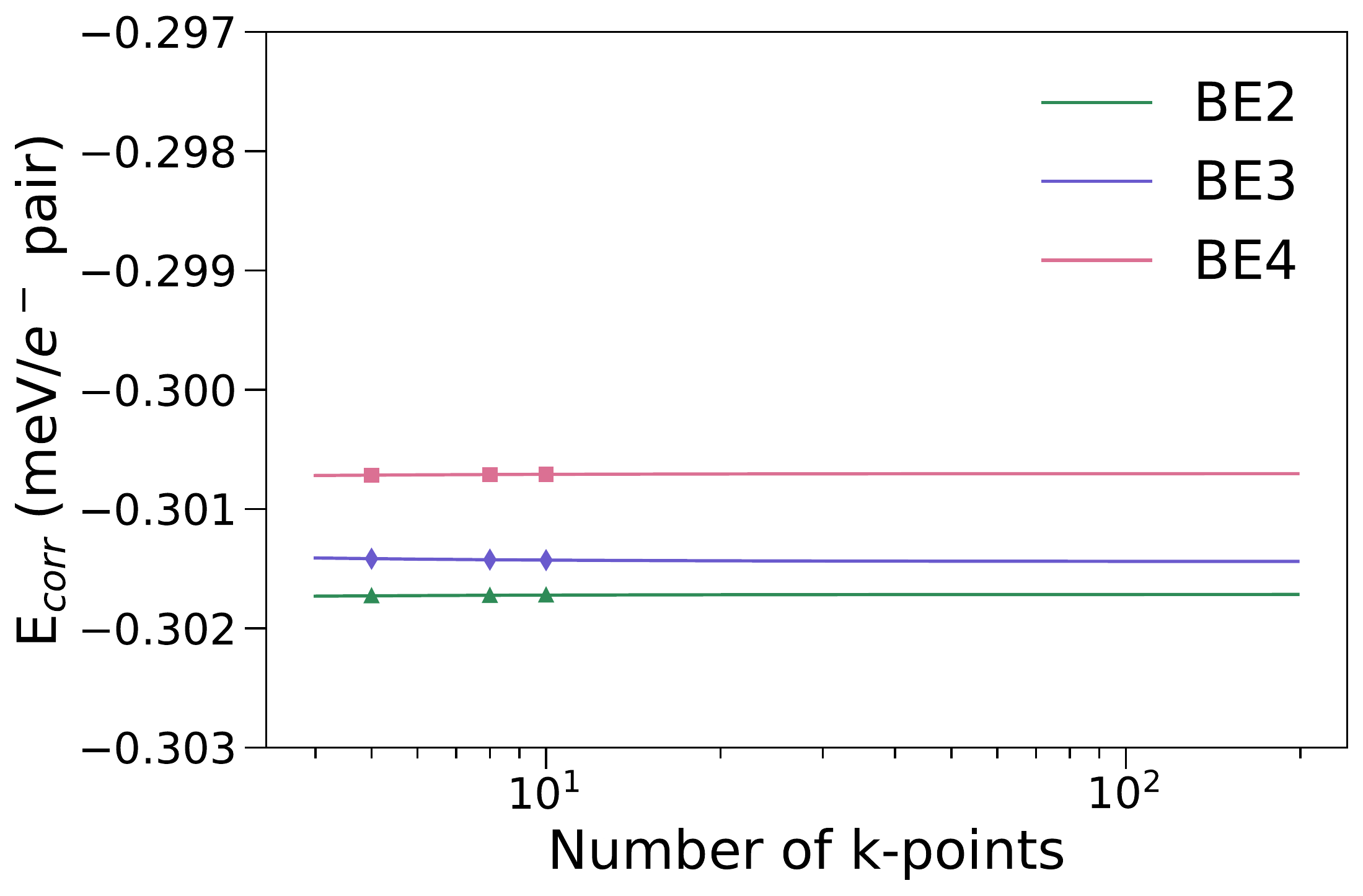}
\end{minipage}
\caption{Convergence of the total electron correlation energy per unit cell to the thermodynamic limit for the modified polymers: (a) NT812, and (b) PM6}\label{Fig:large_conv}
\end{figure}

NT812 and PM6 are two polymer solar cells that exhibit promising photovoltaic performances\cite{https://doi.org/10.1002/sus2.10,pm6photo}. Figure \ref{Fig:large_geom} illustrates the unit cell of the two polymers, which have substantial sizes. For both polymers, even a $\Gamma-$point calculation becomes impossible with the full $k$-CCSD, so the reference $k$-CCSD electron correlation energies are unavailable. In contrast, correlated calculations with the BE$n$ schemes are feasible for relatively large reciprocal space mesh sizes. The computed total electron correlation energies per unit cell as a function of $k$-points are presented in Figure \ref{Fig:large_conv}, rescaled per electron pair. The correlation energies converge to the TDL very quickly with only a few k-points for both polymers. 

Although the reference full $k$-CCSD correlation energies are not available, the BE$n$ correlation energies are very close to each other.
The difference in the correlation energy between BE2 and BE3 is 0.0003 and 0.0005 meV per electron pair for PM6 and NT812, respectively. BE3 and BE4 differ by 0.0007 and 0.0009 meV per electron pair in the correlation energy of PM6 and NT812, respectively.
These differences in the correlation energies between the BE$n$ schemes are well within the observed difference between BE3 and BE4 in the polymers discussed in the previous section.
Further details on the correlation energies for comparison are provided in the Supporting Information. 
With the observed consistency in the differences between the BE$n$ schemes, we estimate that the correlation energies recovered by BE3 and BE4 in the test cases of Section \ref{sec:accuracy} will not change in large polymers, such as NT812 and PM6 polymers.
BE3 and BE4 recover electron correlation energy with an error that falls in the window of $\pm$0.1\% in those test cases. Based on this, we predict the electron correlation energies of the modified PM6 and NT812 donor-acceptor type polymer to be 0.3007$\pm$0.0003 and  0.4563$\pm$0.0005 meV per electron pair, respectively, at the TDL. Herein, the correlated calculation at the TDL for such large periodic systems demonstrates the applicable range of systems using periodic BE.

\section{Conclusions}\label{conclusion}

In this work, we presented periodic bootstrap embedding (BE), a new efficient approach to computing accurate electron correlation energy in periodic systems. As a fragment-based method, the Hilbert space dimension of the individual fragments is significantly reduced compared to the full system. Thus, there is a dramatic reduction in the computational cost. As such, complex periodic systems with large unit cells for which even a $\Gamma$-point computation is impossible can be treated with BE to compute accurate electron correlation energy.

The framework presented in this work provides an interface for utilizing existing non-periodic methods to treat periodic systems. Another important aspect of periodic BE is that the correlated fragment calculation has no explicit dependence on reciprocal space. This allows inexpensive access to the electron correlation energy at the thermodynamic limit, which otherwise requires extremely dense k-point sampling with the correlated method.

Using CCSD as the local fragment correlated solver, periodic BE with BE3 or BE4 fragment schemes yields high accuracies for the total electron correlation energy, with a typical error of around 0.1\%.
With the developments presented in this work, we anticipate BE to be an important tool for computing highly accurate electron correlation energies with periodic boundary conditions.
The most obvious future work involves extending periodic BE to larger basis sets (e.g. double- and triple-zeta basis sets with polarization functions) and to higher dimensional periodic systems (i.e. surfaces and bulk solids). Such work is ongoing in our group.  In the meantime, it is clear that BE provides a promising route to studying large polymers at the periodic boundary conditions using the high-level electron correlation methods.

\section*{Supplementary Information}
Supplementary Information is provided for: (i) Convergence of total electron correlation energies to the TDL for the polymers in Section \ref{sec:accuracy}; (ii) Total electron correlation energies, absolute errors, and difference with increasing fragment sizes at the TDL for all polymers used in this work; (iii) Geometric data including unit cell vectors and cartesian coordinates of all the polymers.

\begin{acknowledgments}
This work was funded by a grant from NSF (NSF CHE-2154938).
\end{acknowledgments}

\bibliographystyle{achemso}
\bibliography{kbe.bib}

\end{document}


\title{Supporting Information for Periodic Bootstrap Embedding}

\author{Oinam Romesh Meitei}
\affiliation{Department of Chemistry, Massachusetts Institute of Technology, Cambridge, Massachusetts 02139, United States}

\author{Troy Van Voorhis}
\email{tvan@mit.edu}
\affiliation{Department of Chemistry, Massachusetts Institute of Technology, Cambridge, Massachusetts 02139, United States}
\maketitle

\section{Convergence to the thermodynamic limit}
\begin{figure}[htb!]
\begin{minipage}{1.0\textwidth}
(a)\includegraphics[scale=0.35]{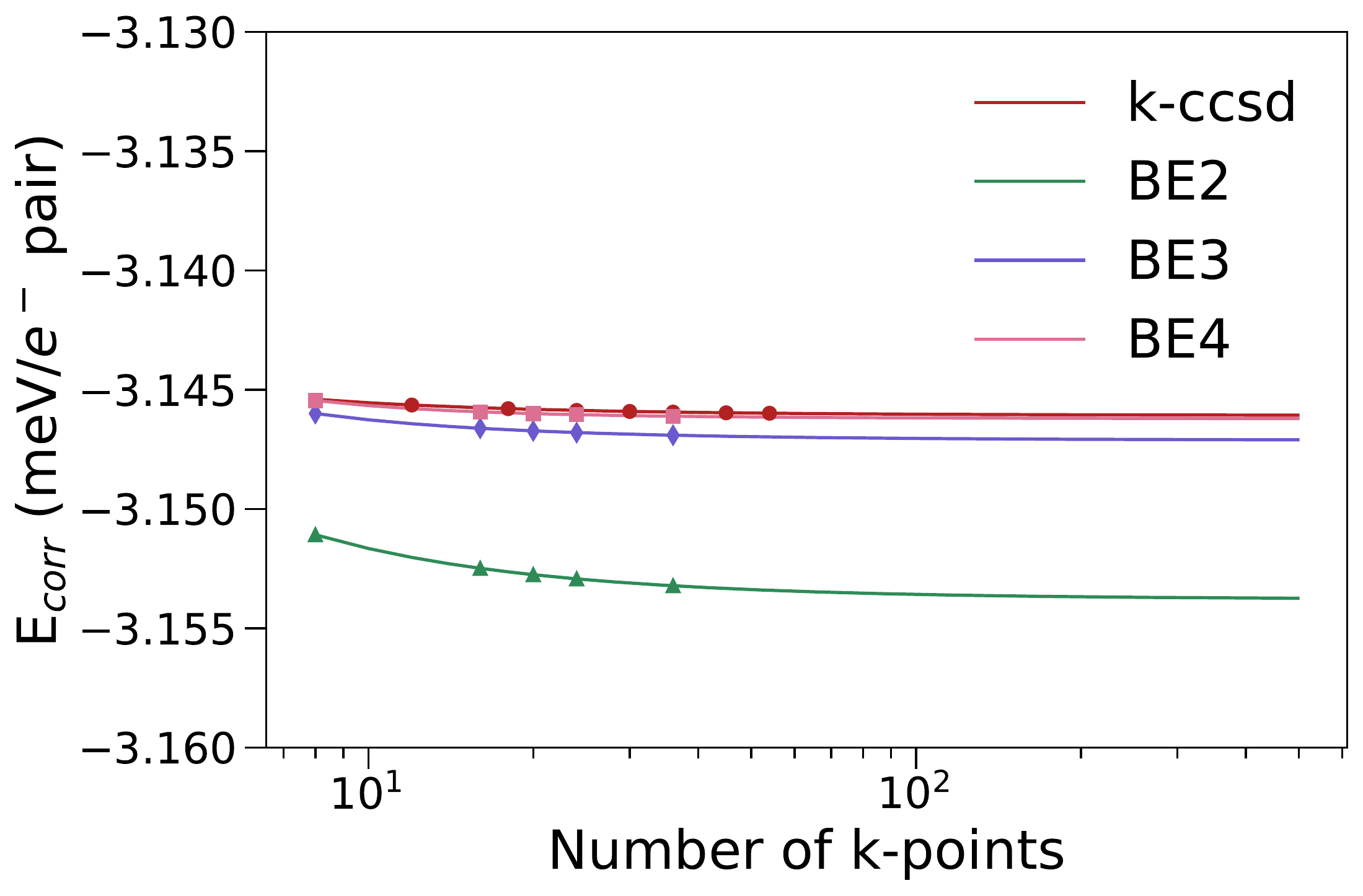}
(b)\includegraphics[scale=0.35]{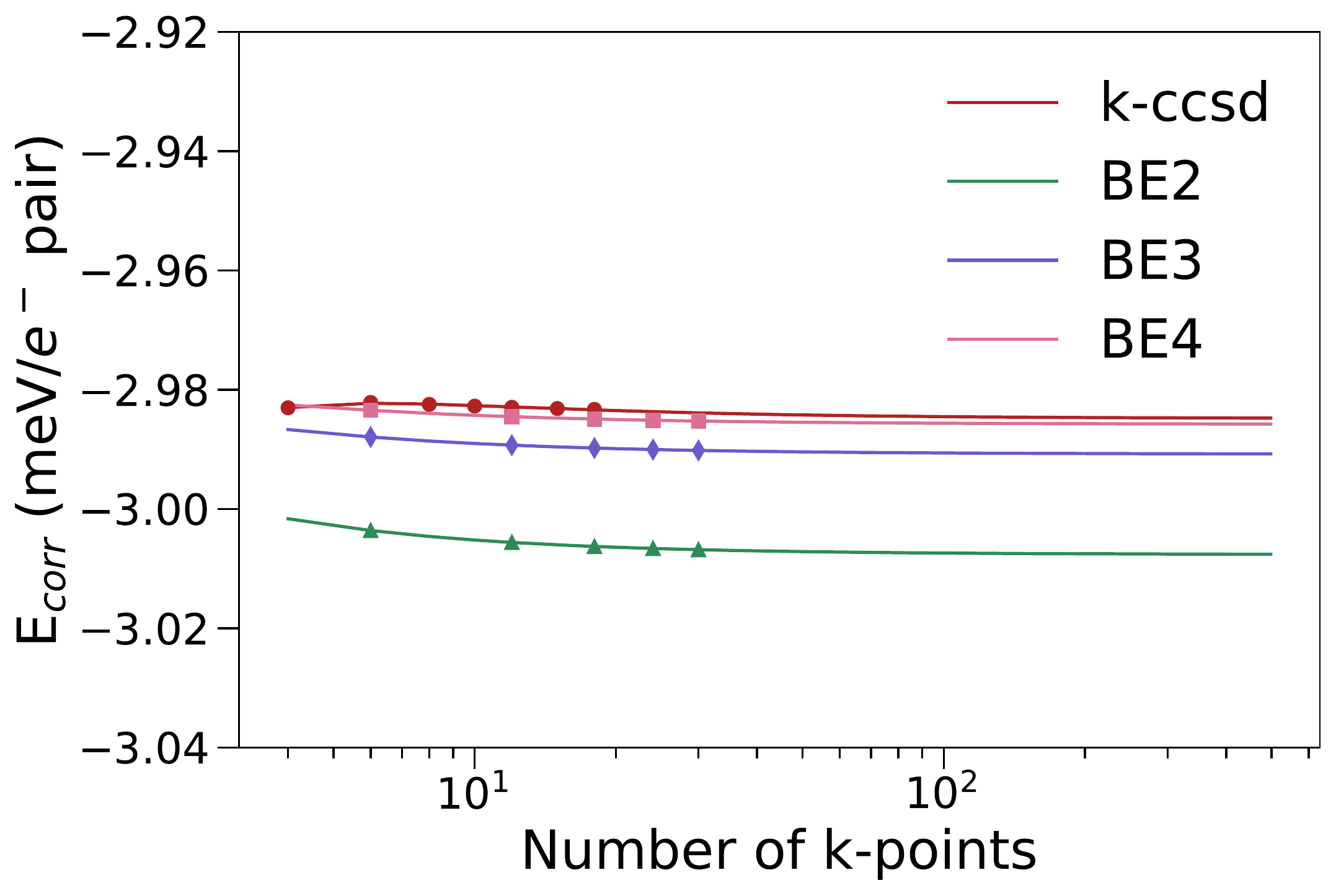}\\
(c)\includegraphics[scale=0.35]{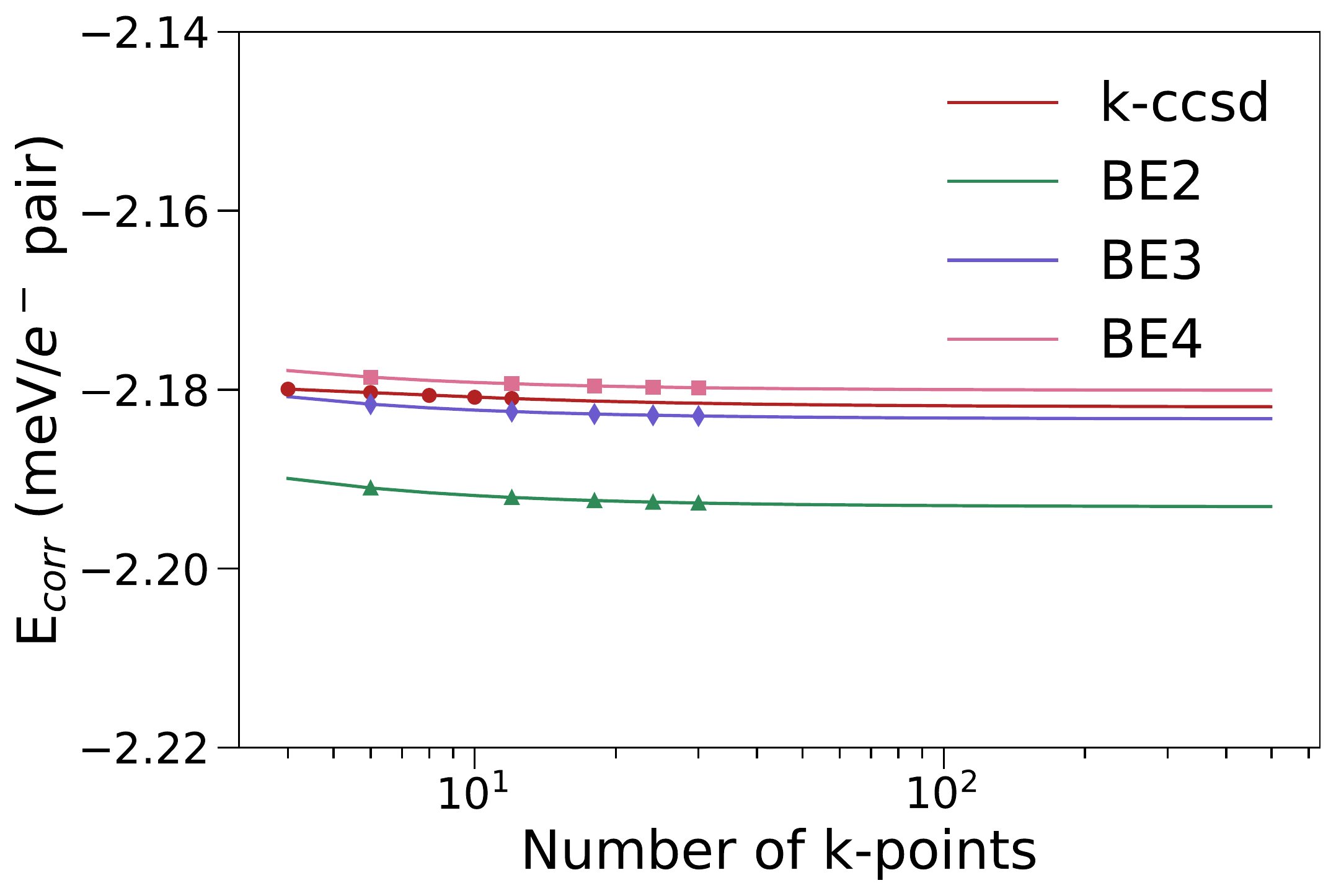}
(d)\includegraphics[scale=0.35]{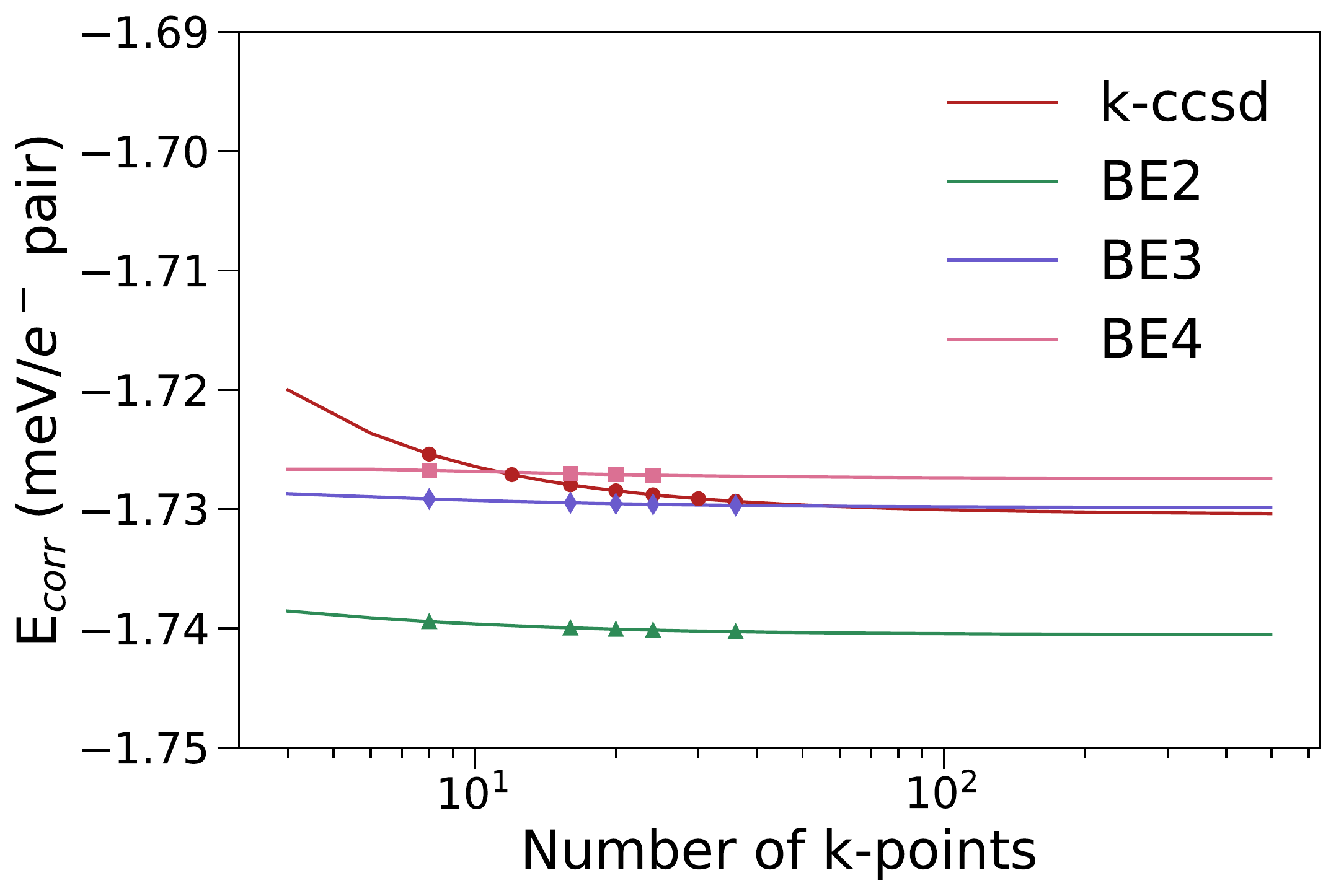}
\end{minipage}
\caption{Convergence of the total electron correlation energy per unit cell to the thermodynamic limit, rescaled per electron pair, for (a) PE, (b) PPP, (c) PPV, and (d) PT.}
\end{figure}

\section{Total electron correlation energies}
\begin{table}[!htb]
  \caption{Total electron correlation energies at the thermodynamic limit (TDL) per unit cell for all the polymers reported in the main text. Units in eV}
  \begin{centering}
    \begin{tabular}{l@{\hspace{3mm}}P{1.5cm}P{1.5cm}P{1.5cm}P{1.5cm}P{1.5cm}P{1.5cm}P{1.5cm}}
      \noalign{\vskip2mm}
    \hline\hline
      \noalign{\vskip2mm}
          & PA & PE & PPP & PPV & PT & PM6 & NT812 \tabularnewline[2mm]
      \hline
      \noalign{\vskip2mm}
      $k$-CCSD         &  -4.0634 & -3.7753 & -11.2827 & -15.3390  & -8.5831 &          &  \tabularnewline[2mm]
      BE2              &  -4.0987 & -3.7845 & -11.3689 & -15.4174  & -8.6332 & -79.4570 &  -58.2207 \tabularnewline[2mm]
      BE3              &  -4.0721 & -3.7765 & -11.3052 & -15.3484  & -8.5803 & -79.3843 &  -58.1559 \tabularnewline[2mm]
      BE4              &  -4.0662 & -3.7755 & -11.2863 & -15.3260  & -8.5682 & -79.1903 &  -58.0415 \tabularnewline[2mm]
      \hline\hline
    \end{tabular}
    \par\end{centering}
\end{table}

\begin{table}[!htb]
  \caption{Absolute error in the BE$n$ total correlation energies compared to $k$-CCSD at the TDL, rescaled per electron pair. Units in meV.}
  \begin{centering}
    \begin{tabular}{l@{\hspace{3mm}}P{2cm}P{2cm}P{2cm}}
      \noalign{\vskip2mm}
    \hline\hline
      \noalign{\vskip2mm}
          & BE2 & BE3 & BE4  \tabularnewline[2mm]
      \hline
      \noalign{\vskip2mm}
      PA  & 0.0389 & 0.0096 & 0.0031 \tabularnewline[2mm]
      PE  & 0.0077 & 0.0010 & 0.0001 \tabularnewline[2mm]
      PPP & 0.0228 & 0.0060 & 0.0010 \tabularnewline[2mm]
      PPV & 0.0112 & 0.0014 & 0.0018 \tabularnewline[2mm]
      PT  & 0.0101 & 0.0006 & 0.0030 \tabularnewline[2mm]
      \hline\hline
    \end{tabular}
    \par\end{centering}
\end{table}

\begin{table}[!htb]
  \caption{Difference in the total correlation energies with increasing fragment sizes in BE at the thermodynamic limit (TDL), rescaled per electron pair. Units in meV.}
  \begin{centering}
    \begin{tabular}{l@{\hspace{3mm}}P{3cm}P{3cm}}
      \noalign{\vskip2mm}
    \hline\hline
      \noalign{\vskip2mm}
          & $E^{corr}_{BE2} - E^{corr}_{BE3}$  & $E^{corr}_{BE3} - E^{corr}_{BE4}$   \tabularnewline[2mm]
      \hline
      \noalign{\vskip2mm}
      PA         &  -0.0067 & -0.0009  \tabularnewline[2mm]
      PE         &  -0.0293 & -0.0065  \tabularnewline[2mm]
      PPP        &  -0.0168 & -0.0050  \tabularnewline[2mm]
      PPV        &  -0.0098 & -0.0024  \tabularnewline[2mm]
      PT         &  -0.0107 & -0.0024  \tabularnewline[2mm]
      PM6        &  -0.0003 & -0.0007  \tabularnewline[2mm]
      NT812      &  -0.0005 & -0.0009  \tabularnewline[2mm]
      \hline\hline
    \end{tabular}
    \par\end{centering}
\end{table}

\section{Geometric data}
The unit cell vectors and cartesian coordinates for the isolated polymer chains used in main text are provided below. The lattice vectors are denoted as a$i$ and units are in Angstrom. As mentioned in the main text, the structures of PM6 and NT-812 are modified by removing the side alkane chains and replacing it by a hydrogen atom.

\begin{verbatim}
Polyacetylene (PA) chain

a1       8.0    0.0    0.0  
a2       0.0    8.0    0.0  
a3       0.0    0.0    2.455
H        1.42856     0.0       -0.58617    
C        0.34156     0.0       -0.58799    
H       -1.42856     0.0        0.58617    
C       -0.34156     0.0        0.58799    
\end{verbatim}

\begin{verbatim}
Polyethylene (PE) chain

a1       8.0    0.0    0.0  
a2       0.0    8.0    0.0  
a3       0.0    0.0    2.6985
H       -1.17242    -0.70545    0.50328    
H       -0.91008     1.04871    0.50796    
C       -0.38115     0.07288    0.51740    
H        1.11637     0.70473    1.88084    
H        0.85404    -1.04943    1.87616    
C        0.32511    -0.07360    1.86671
\end{verbatim}

\begin{verbatim}
Poly(p-phenylene) (PPP)

a1      20.0    0.0    0.0  
a2       0.0   20.0    0.0  
a3       0.0    0.0    4.310
C        0.00000     0.0        0.00000                 
C        1.20432     0.0        0.72362                 
C        1.20674     0.0        2.11262                 
C        0.00000     0.0        2.84045                 
C       -1.19936     0.0        2.11682                 
C       -1.20178     0.0        0.72782                 
H        2.16146     0.0        0.17546                 
H        2.17051     0.0        2.64905                 
H       -2.15650     0.0        2.66498                 
H       -2.16555     0.0        0.19140   
\end{verbatim}

\begin{verbatim}
Poly(p-phenylene vinylene) (PPV)

a1      20.0    0.0    0.0  
a2       0.0   20.0    0.0  
a3       0.0    0.0    6.650
C        0.00000     0.0        0.00000                 
C       -1.31103     0.0       -0.52969                 
C       -1.54167     0.0       -1.89332                 
C       -0.47776     0.0       -2.82471                 
C        0.83505     0.0       -2.30757                 
C        1.06569     0.0       -0.94394                 
H       -2.17743     0.0        0.15454                 
H       -2.57843     0.0       -2.27272                 
H        1.69103     0.0       -3.00321                 
H        2.10348     0.0       -0.57031                 
C        0.23443     0.0        1.42382                 
H        1.30136     0.0        1.72272                 
C       -0.71695     0.0        2.39705                 
H       -1.76236     0.0        2.02993    
\end{verbatim}

\begin{verbatim}
Polythiophene (PT)

a1      30.0    0.0    0.0  
a2       0.0   30.0    0.0  
a3       0.0    0.0    7.8562
C        3.74360     5.55710    7.14890
C        3.18510     4.41510    6.58860
C        3.18510     4.41510    5.17210
C        3.74360     5.55710    4.61180
H        2.79260     3.59960    4.57700
H        2.79260     3.59960    7.18370
S        3.39270     4.78350    9.80840
S        4.27710     6.66240    5.88040
C        3.92620     5.88880   11.07700
C        4.48470     7.03070   10.51670
C        4.48470     7.03070    9.10020
C        3.92620     5.88870    8.53990
H        4.87720     7.84630    8.50510
H        4.87720     7.84630   11.11180
\end{verbatim}
\pagebreak

\begin{verbatim}
PM6 (modified)

a1       20.006  0.0    0.0  
a2        0.0   40.0    0.0  
a3        0.0    0.0   40.0
C         4.5994667910        0.3365357991       -0.3185613144
C         4.2988542490       -1.0214163249       -0.1162176619
C         2.9394647419       -1.3366169862       -0.2417524961
C         2.1664500899       -0.2155589095       -0.5556133509
S         3.1330092107        1.2093363282       -0.6769264376
H         2.5227090782       -2.3228716152       -0.0600992171
H         5.0500369436       -1.7556468352        0.1621732092
C         8.4732398857        1.0187980170       -0.0018739590
C         7.8584651660        2.2784822525       -0.0383796686
C         6.4201360458        2.2173799307       -0.1966529859
C         5.9195366542        0.9056665032       -0.2381094484
S         7.2474473728       -0.2162035628       -0.1518454148
C         5.5885983049        3.4208084219       -0.2693454815
C         8.5756199494        3.5490077765        0.0968963421
C         9.8351921210        0.5499898584        0.0805050622
S        11.3066165489        1.5020775583        0.0047580904
C        12.3074841255        0.0835683858        0.0063547083
C        11.5330087999       -1.0755439443        0.1046576281
C        10.1647694259       -0.8112139296        0.1333703928
H        11.9437872411       -2.0810032086        0.1582424905
H         9.4224713535       -1.6036513388        0.1835692533
O         4.3550736493        3.4042368690       -0.4476140270
C         6.3157663828        4.6900599256       -0.1367267284
C         7.7367533081        4.7546877868        0.0218019612
O         9.8054033931        3.6250118636        0.2726239457
C         5.7328942294        5.9403230229       -0.1795129771
C         8.1934782121        6.0552307521        0.0956380043
S         6.9060348412        7.1728364480       -0.0248165524
H         4.6783752143        6.1690173110       -0.2836566862
H         9.2191251648        6.3890538644        0.1886027014
C         0.7343062117       -0.1073912247       -0.6543252227
S        -0.0202866861        1.4746015295       -0.6058393421
C        -1.6149613559        0.7816409512       -0.5629663684
C        -1.5428204673       -0.6458113316       -0.6396962704
C        -0.2024049919       -1.1199912951       -0.6888239182
H         0.0545824242       -2.1748829464       -0.7278775068
C        -2.8200072799        1.4976671084       -0.4642409547
C        -4.0021753060        0.7122075448       -0.3414063410
C        -3.9212447693       -0.7118404055       -0.4521686050
C        -2.7288367149       -1.4169726566       -0.6393018110
C        -5.3493143758        1.1417610268       -0.1297656800
H        -5.6263475174        2.1790915229        0.0428233175
C        -6.2632490883        0.1036688557       -0.1240270339
S        -5.4847406678       -1.4479222406       -0.3703256943
C        -2.7265532117       -2.8666659864       -0.8605415526
C        -3.2045450489       -3.8711387977       -0.0380804001
H        -3.5884912995       -3.7028815947        0.9627872048
C        -3.1319359334       -5.1407722003       -0.6703270357
C        -2.6020675202       -5.1208217691       -1.9380585552
H        -2.4566158531       -5.9432220087       -2.6266867241
S        -2.1925218145       -3.5176160038       -2.3790515475
F        -3.5470942232       -6.2768457737       -0.0647735157
C        -2.8404217136        2.9613825416       -0.5595167283
S        -3.9410927248        3.7628782589       -1.6475406619
C        -3.2896568342        5.3141495065       -1.3242533652
H        -3.6710324700        6.2001944722       -1.8148655712
C        -2.2650605562        5.2098271639       -0.4135188976
C        -1.9891475748        3.8893895387        0.0214293988
F        -1.5677730859        6.2739921108        0.0525513659
H        -1.2126570682        3.6561100876        0.7421291845    
\end{verbatim}

\begin{verbatim}
NT-812 (modified)

a1      19.724  0.0    0.0  
a2       0.0   40.0    0.0  
a3       0.0    0.0   40.0
C        -0.0395309979        1.3587114386       -0.2103535195
C        -0.4773902138        2.6781170748       -0.1164474255
C        -1.8761497365        2.8054071262       -0.1225534973
C        -3.9706735134        1.3340351962       -0.2498745246
C        -4.4884393364        0.0429574759       -0.2165743264
C        -2.5478976706        1.5890474843       -0.2188135752
S        -1.3958396577        0.2800999823       -0.3063721262
C        -5.8722318386       -0.2423998850       -0.2346265192
C        -6.3438204910       -1.5965284218       -0.2151088720
C        -7.7519066907       -1.9115502596       -0.2064297929
C        -8.7466730301       -0.8616345180       -0.1986521475
C        -8.2504503789        0.4427916451       -0.2513026425
C        -6.8701250732        0.7619755010       -0.2756725423
C        -6.3765451739        2.1125843949       -0.3359490931
H        -8.9449088307        1.2810893390       -0.2603699783
N        -7.9807038634       -3.2362286188       -0.2099145486
S        -6.5318039006       -3.9715024273       -0.2123188018
N        -5.5491433878       -2.6774039387       -0.2161573229
C        -4.9535259529        2.3950339986       -0.3294524145
N        -7.1443854004        3.2118529466       -0.4264581067
N        -4.6853040819        3.7081841379       -0.4215103180
S        -6.1208232708        4.4770990660       -0.4970517715
C         1.3143522621        0.8854877573       -0.2179177103
C         1.7859616535       -0.4250308928       -0.2279452954
S         2.6281108607        2.0528533260       -0.1990359405
C         3.1945706194       -0.4681006043       -0.2147273731
C         3.8103749433        0.7954706035       -0.1929408211
S         4.3806207815       -1.7237997336       -0.1899764201
C         5.2168136777        0.7548721952       -0.1351871533
C         5.6904413421       -0.5551055524       -0.1293480029
C         7.0503486892       -1.0104942626       -0.0966778908
S         8.3688888673        0.0969079517       -0.3214771786
C         9.5630922061       -1.1562377842       -0.1262862516
C         8.9341048028       -2.3817818747        0.0919396683
C         7.5316885299       -2.3033071346        0.1058188921
H         0.2071298849        3.5194060060       -0.0445025065
H        -3.8190047369       -0.8146595850       -0.1608710398
H         1.1359843218       -1.2929839052       -0.2402338141
H         9.5009692484       -3.2940005780        0.2407026201
H         6.8846860098       -3.1616894137        0.2711248899
H         5.8646866096        1.6229850398       -0.0791417035
H        -2.4071224366        3.7469130816       -0.0592163341  
\end{verbatim}

\vspace{-2em}